\documentclass[11pt,a4paper]{article}
\usepackage{jcappub}

\newcommand{\be}{\begin{eqnarray}}
\newcommand{\bee}{\begin{enumerate}}
\newcommand{\bit}{\begin{itemize}}

\def\bkt#1{\left(#1\right)} % normal brackets
\def\bkts#1{\left[#1\right]} % square brackets
\def\bkta#1{\langle#1\rangle} % angle brackets

\def\diff#1#2{{d{#1}\over d{#2}}}
\def\ee{\end{eqnarray}}
\newcommand{\eee}{\end{enumerate}}
\def\eg{\textit{e.g.} }
\newcommand{\eit}{\end{itemize}}
\def\etal{\textit{et al.}} % for end of sentence 
\def\fnl{\ensuremath{f_{\mbox{\scriptsize NL}}}}
\def\ff{\phantom{.}}

\def\half{\ensuremath{{1\over2}}}
\def\ie{\textit{i.e.}}
\def\iee{\textit{i.e. }}
\newcommand{\ii}{\textit}

\def\lab{\label}

\newcommand{\mb}{\mathbf}
\def\mc#1{\mathcal{#1}}
\def\ms#1{\mathscr{#1}}
\newcommand{\mmm}{\medskip}

\newcommand{\no}{\noindent}
\newcommand{\nn}{\nonumber}

\def\pdiff#1#2{{\partial{#1}\over \partial{#2}}}

\def\re#1{(\ref{#1})}

\def\sub#1{_{\mbox{\scriptsize{#1}}}}
\def\sun{\odot}
\def\super#1{^{\mbox{\scriptsize{#1}}}}

\def\xHI{\ensuremath{\bar{x}\sub{HI}} }
\usepackage{amsmath}
\usepackage{amsfonts}
\usepackage{amssymb}
\usepackage{graphicx}
\usepackage{epsfig}
\usepackage{hyperref}
\usepackage{mathrsfs}

\usepackage{bm}

\begin{document}

\title{The 21cm power spectrum and the shapes of non-Gaussianity}

\author{Sirichai Chongchitnan}

\affiliation{School of Engineering, Computing and Applied Mathematics, 
University of Abertay Dundee, Bell St., Dundee, DD1 1HG, Scotland.}

\emailAdd{s.chongchitnan@abertay.ac.uk}

\abstract{We consider how measurements of the 21cm radiation from the epoch of reionization ($z=8-12$) can constrain the amplitudes of various `shapes' of primordial non-Gaussianity. The limits on these shapes, each parametrized by the non-linear parameter $\fnl$, can reveal whether the physics of inflation is more complex than the standard single-field, slow-roll scenario. In this work, we quantify the effects of the well-known local, equilateral, orthogonal and folded types of non-Gaussianities on the 21cm power spectrum, which is expected to be measured by upcoming radio arrays such as the Square-Kilometre Array (SKA). We also assess the prospects of the SKA in constraining these non-Gaussianities, and found constraints that are comparable with those from cosmic-microwave-background experiments such as \ii{Planck}. We show that the limits on various $\fnl$ can be tightened to $\mc{O}(1)$ using a radio array with a futuristic but realistic set of specifications.}

\maketitle

\section{Introduction}

The statistical nature of primordial density perturbations can reveal much about the physics of the early Universe. If the simplest theory of single-field slow-roll  inflation is correct, the perturbations should obey an essentially Gaussian statistics \citep{maldacena}. Any detection of primordial non-Gaussianity would hint at novel physics during inflation, or perhaps even an alternative theory of the early Universe (see \cite{chen,bartolo} for reviews).

Much effort in the search for primordial non-Gaussianity has been directed towards constraining the `local' non-Gaussianity parameter, $\fnl$, which occurs in the expansion of the nonlinear Newtonian potential, $\Phi$, as a series of Gaussian random field, $\phi$,
\be \Phi = \phi + \fnl(\phi^2-\bkta{\phi^2})+\ldots \ee
The most stringent constraint on $\fnl$ comes from the cosmic microwave background (CMB) anisotropies which give $\fnl=32\pm42$ $(2\sigma)$ \cite{komatsu}. 

In addition, there may be other kinds of non-Gaussianity, including the so-called equilateral, orthogonal and folded non-Gaussianities which, as we explain in more detail in the next section, are signatures of specific physics in the inflationary era. We refer to these varieties as \ii{shapes} of non-Gaussianity, as they are associated with different 3-point correlation functions, $B(\mb{k}_1,\mb{k}_2,\mb{k}_3)$, which peak when the Fourier wavevector, $\mb{k}_i$, of the perturbations form different triangular configurations \cite{creminelli,meerburg,senatore}. Current constraints on these shapes are much weaker than that of the local model. At $2\sigma$, the constraints from WMAP are $\fnl\super{equil}=26\pm240$, $\fnl\super{orth}=-202\pm208$ \cite{komatsu} (there is no published constraint on $\fnl\super{folded}$ at present). These constraints are expected to be improved by at least an order of magnitude with the highly anticipated results from the \ii{Planck} satellite\footnote{\texttt{www.rssd.esa.int/planck}}, as well as  constraints from various upcoming large-scale-structure (LSS) surveys (see \eg \cite{desjacques} for a review).

More recently, it has been shown that non-Gaussianity could leave a detectable signature in the 21cm radiation due to spin-flip transitions of neutral hydrogen during the epoch of reionization or earlier during the cosmic Dark Ages \cite{cooray,joudaki,pillepich,tashiro,tashiro1,me}. In particular, Joudaki \etal \cite{joudaki} showed that there are good prospects of constraining local $\fnl$ with the Square-Kilometer Array (SKA)\footnote{\texttt{www.skatelescope.org}} to within $|\fnl|\lesssim10$ using measurements of the 21cm power spectrum from redshift $z\sim7$. In this work, we extend this pursuit to other shapes of non-Gaussianity. In particular, we investigate the prospects of the SKA in constraining $\fnl$ associated with different non-Gaussianity shapes, in comparison with the CMB or LSS constraints. 

Throughout this work we assume in our fiducial model the following cosmological parameters \cite{lahavliddle}: present density parameter for matter($\Omega_m=0.276$), baryon ($\Omega_b=0.046$), radiation ($\Omega_r=5\times10^{-5}$), dark energy ($\Omega\sub{DE}=1-\Omega_m-\Omega_r$); scalar spectral index, $n_s=0.96$, and normalization of matter power spectrum, $\sigma_8=0.81$.   

%Main questions to tackle: 
%a) How do different shapes of non-Gaussianity manifest in the 21cm power spectrum (from the IGM)?
%b) Can the SKA discern / distinguish between these shapes? (issue of noise)
%c) Is it a a better discriminant compared with the CMB? or LSS?
%d) How sensitive are the conclusions to gas physics?

\section{Shapes of non-Gaussianity}

The bispectrum, $B_\Phi$, is defined as the three-point correlation function of the Newtonian potential, $\Phi$. In this work, we shall consider four common forms of the bispectrum described below.

\mmm

\no 1. \ii{The local model} is the dominant type of non-Gaussianity in single field inflation and curvaton models \cite{gangui,verdewang,komatsuspergel}. The bispectrum is given by 
\be B\super{loc}(\mb{k}_1,\mb{k}_2,\mb{k}_3)&=& 2\fnl\super{loc}\mathcal{F}\super{loc},\\
\mathcal{F}\super{loc}&=&P_1P_2 + 2 \mbox{ perm.}
\ee
where $P_i=P_\Phi(k_i)\propto k_i^{n_s-4}$ is the primordial power spectrum of $\Phi$. In wavevector space, the bispectrum peaks when the vectors $\mb{k}_1, \mb{k}_2, \mb{k}_3$ form a `squeezed' triangle in which $k_1\approx k_2\gg k_3$.

\mmm

\no 2. \ii{The equilateral model} of non-Gaussianity is dominant in inflation models with higher-derivative terms in the Lagrangian. The bispectrum peaks when $\mb{k}_1,\mb{k}_2,\mb{k}_3$ form an equilateral triangle \cite{creminelli}. The precise expression for the bispectrum is model-dependent and may take a complicated form, but for computational purposes they can be approximated by a `template' which comprises factorisable pieces of the power spectrum:
\be B\super{eql}(\mb{k}_1,\mb{k}_2,\mb{k}_3)&=&6\fnl\super{eql}\bkt{-\mathcal{F}\super{loc}-2\mathcal{F}^{A}+\mathcal{F}^{B}},\\
\mathcal{F}\super{A}&=& \bkt{P_1P_2P_3}^{2/3},\\
\mathcal{F}\super{B}&=& P_1^{1/3}P_2^{2/3}P_3+ 5 \mbox{ perm.}
\ee
(see \cite{wagner} for numerical comparison between the templates and physical shapes.)
\mmm

\no 3. \ii{The orthogonal model} was constructed to be orthogonal to the equilateral shape with respect to an inner product of bispectra as defined in \cite{senatore}. It also occurs in models with higher derivatives in the Lagrangian. The template for the bispectrum is
\be B\super{orth}(\mb{k}_1,\mb{k}_2,\mb{k}_3)&=&6\fnl\super{orth}\bkt{-3\mathcal{F}\super{loc}-8\mathcal{F}^{A}+3\mathcal{F}^{B}},\ee
which peaks on equilateral and folded configurations (see below). 

\mmm

\no 4. \ii{The folded model} of non-Gaussianity is prevalent in inflation models in which  the initial Bunch-Davies vaccum state is modified \cite{meerburg}. The template for the bispectrum is
\be B\super{fol}(\mb{k}_1,\mb{k}_2,\mb{k}_3)={1\over 2}\bkt{B\super{eql}-B\super{orth}},\ee
peaking on a flattened triangle configuration with $k_1=2k_2=2k_3$.

\section{A model of reionization}

We are looking for signatures of the various non-Gaussianity shapes in the 21cm signal from the epoch of reionization $(z\sim8-12)$. %The main sources of 21cm radiation we are interested in are clouds of hydrogen in the epoch of reionization . 
The intensity of the 21cm radiation depends on the local ionized fraction, $x_i$. We define the ionized-fraction contrast as
\be \delta_x\equiv {x_i - \bar{x}_i\over \bar{x}_i}, \ee
where $\bar{x}_i$ is the mean ionized fraction. Both $x_i$ and $\bar{x}_i$ depend on the model of reionization. 

In this work, we appeal to the reionization model of Alvarez \etal \ff\cite{alvarez}, in which the ionized fraction $x_i$ is a function of the local overdensity, $\delta$, and redshift, and satisfies
\be \ln[1-{x_i}(\delta,z)]=-\zeta(\delta,z) f\sub{coll} (\delta,z,M),\lab{xx}\ee
where $\zeta$ is the reionization efficiency and  $f\sub{coll}$ is the collapse fraction in a region associated with a top-hat smoothing function with smoothing scale $R$ (or smoothing mass $M=4\pi R^3\rho_0/3$). Alvarez \etal \ff noted that this simple parametrization reproduces the behaviour expected when the number of ionizing photons is large ($\zeta f\sub{coll}\gg1$) and small compared with atoms (see their Appendix for detail). A particularly simple form of $\zeta$ can be obtained in the  ``Str\"omgren" limit in which a photon is emitted for every HI atom formed upon recombination. In this case, it can be shown that
\be\zeta(\delta)={\zeta_0\over 1+\delta}\approx\zeta_0(1-\delta),\ee with $\zeta_0$ a constant and $|\delta|\ll1$, which holds in our application.

The collapse fraction depends on the halo formation model. For example, in the extended Press-Schechter theory \cite{bcek,lacey}, $f\sub{coll}$ depends on the difference between the local overdensity $\delta(z)$ and the critical collapse overdensity $\delta_c(z)$\footnote{We shall make a strict distinction between $\delta_c$ and $\delta_c(z)$: $\delta_c(z)= \delta_c/ D(z)$ and $\delta(z)= \delta/ D(z)$, where $\delta_c$ and $\delta$ are overdensities extrapolated to $z=0$ and $D(z)$ is the growth function.
}.
%be f\sub{coll}(z)= {1\over \bar{\rho}_0}\int_{M\sub{min}}^{\infty} \diff{n}{M} dM\ee
%where $\rho_0$ is the matter density at present day. The Press-Schechter formalism gives % DERIVE THIS.. excursion set.. yes. not easy.. Zentner review eq 31
\be f\sub{coll}(\delta,z,M)= \mbox{erfc}\bkt{\delta_c(z) - \delta(z) \over \sqrt{2(\sigma^2(M\sub{min})-\sigma^2(M))}},\lab{fcoll}\ee
where $M\sub{min}$ is the minimum mass for collapse, which we take to correspond to the virial temperature of $10^4$ K (at which point  atomic hydrogen line cooling is efficient) 
\be M\sub{min}= 2.77\times 10^7 h^{-1}M_\sun \bkt{z+1 \over 10}^{-3/2}\bkt{h^2\Omega_m \over 0.15}^{-1/2}. \ee

Following Tashiro \& Ho, we approximate the relation between the matter density contrast, $\delta$, and the ionized-fraction contrast $\delta_x$ by 
\be \delta_x = b_x \delta,\lab{fracion}\ee
where $b_x$ is referred to as the bias parameter for the ionized fraction. We can obtain a relation between $b_x$ and the usual halo bias, $b$, by expanding \re{fcoll} about small $|\delta|$ and equating the result with $f\sub{coll}=f\sub{coll}(0)(1+(\bar{b}-1)\delta)$, yielding
\be b_x = {\bar{x}_i -1 \over \bar{x}_i} \ln(1-\bar{x}_i)(\bar{b}-2),\lab{alva}\ee
where $\bar{b}$ is the average bias of dark matter halos with mass above $m\sub{min}$
\be\bar{b}= 1+ \sqrt{2\over\pi} {\nu\sub{min}e^{-\nu\sub{min}^2/2} \over \delta_c\ff\mbox{erfc}(\nu\sub{min}/\sqrt{2})},\ee
with $\nu\equiv \delta_c(z)/\sigma$ and $\nu\sub{min}\equiv \delta_c(z)/\sigma\sub{min}$. The expression for the mean ionized fraction can be obtained by setting $\delta=0$ in \re{xx},
\be \ln(1-\bar{x}_i)=-\zeta_0\ff\mbox{erfc}\bkt{\nu\sub{min}/\sqrt{2}}
.\lab{xbar}\ee
%where $\zeta_0$ is constant, which we set to be $50$. (???)

%ignore non-Gaussianity at this point as it has a small effect [really?].

\no In summary, the ionized-fraction bias, $b_x$, can be calculated once a history of reionization is known. 

As pointed out in \citep{alvarez}, the average halo bias $\bar{b}$ can also be calculated by averaging the mass-dependent Eulerian bias over the multiplicity function $f(\nu)$ (see \citep{mowhite} for details).
\be \bar{b} &=&  {\int_{\nu\sub{min}}^\infty f\sub{PS}(\nu) b\sub{PS}(\nu) d\ln\nu\over   \int_{\nu\sub{min}}^\infty f\sub{PS}(\nu) d\ln\nu}, \lab{ioi}\\
b\sub{PS}&=& 1+ {\nu^2-1 \over \delta_c},\lab{bLps} \\
 f\sub{PS}(\nu)&=& \nu e^{-\nu^2/2}.\ee
 
This gives us a way to extend the bias calculation to other collapse formalisms. We outline the calculation using the Sheth-Tormen formalism \cite{sheth} in the Appendix, and comment on the results in section \ref{fishsec}

In the next section we shall see how the various non-Gaussianity shapes manifest in the ionized-fraction bias.

%[compare with the $\delta_c\rightarrow q\delta_c$ replacement in $f\sub{coll}\super{PS}$?]

\section{The 21cm power spectrum in the presence of non-Gaussianities}

%if can measure bispectrum then see pillepich. here we focus on whether the 2-point information is sufficient.

The intensity of the observed 21cm signal can be measured by the \ii{differential} brightness temperature, $\delta T_b$ (the difference between the 21cm and CMB temperatures). In the optically thin limit, it can be shown that the mean brightness temperature is given by (see \eg \citep{zaldarriaga})
\be \overline{\delta T_b}(z)\approx 23.88 \mbox{ mK }\xHI \bkt{1-{T\sub{CMB}\over T_S}} {\Omega_b h^2\over 0.02} \bkt{{0.15\over \Omega_m h^2}{1+z \over 10}}^{1/2},\ee
where \xHI is the average neutral fraction of hydrogen ($\xHI+\bar{x}_i=1$).

%Define the matter-density power spectrum as $\bkta{\delta(\mb{k}),\delta(\mb{k^\pr})}= (2\pi)^3\delta^3(\mb{k}-\mb{k^\pr}) P_{\delta\delta}(\mb{k})$

Assuming the fluctuations are linear, one can show that the 21cm power spectrum can be expanded as \cite{bharadwaj,barkana}
\be P_{\delta T} (\mb{k})=\ms{P}_{\delta\delta}-2\ms{P}_{x\delta}+\ms{P}_{xx} + 2\bkt{\ms{P}_{\delta\delta}- \ms{P}_{x\delta}}\mu^2+\ms{P}_{\delta\delta}\mu^4,\lab{power}\ee
where $\mu$ is the cosine of the angle between wavevector $\mb{k}$ and the line-of-sight vector $\mb{\hat n}$.
% explain where are the rest of the mu terms. where does this come from????
Here we define $\ms{P}_{\delta\delta}\equiv \overline{\delta T}_b^2\xHI^2 P_L,$  where $P_L(k)$ is the linear power spectrum $\propto \mc{M}^2(k)k^{n_s-4}$, and
\be\mc{M}(k)\equiv {2k^2 T(k)D(z)\over 3H_0^2\Omega_m},\ee 
%Whilst this coefficient can be evaluated once the cosmological parameters are defined,  the remaining coefficients depend on the details of  reionisation and are more difficult to calculate analytically. It is common to consolidate the astrophysical effects into a single `bias' factor, $b_x(k)$, which quantifies clustering of HI in relation to that of the underlying dark matter. 
where $T(k)$ is the transfer function and $D(z)$ the growth factor. The remaining components of \re{power} are defined by
\be \ms{P}_{x\delta}&=& b_x \ms{P}_{\delta\delta},\\
 \ms{P}_{xx}&=& b_x^2 \ms{P}_{\delta\delta},\ee
which are the density-ionization cross-correlation and the ionization autocorrelation respectively. These expressions reduce the 21cm power spectrum to
\be P_{\delta T} (\mb{k})=\ms{P}_{\delta\delta}\bkt{1-b_x+\mu^2}^2.\lab{shortcut}\ee
Non-Gaussianity affects the 21cm power spectrum by introducing a characteristic scale dependence in the bias $b_x$. This is analogous to the non-Gaussian effects on the halo bias, which has already allowed a number of cluster surveys to place limits on non-Gaussianity, particularly the local type (see \eg \citep{dalal,desjacques} and references therein). Calculations of the halo bias in the presence of other shapes of non-Gaussianity were explored by Matsubara \cite{matsubara} (amongst others \cite{desjacques,verde,wagner}). The calculations in this work are based on the analytic results of \cite{matsubara}.

Writing $b\super{NG} = b\super{Gaussian}+\Delta b$, Matsubara showed using his theory of ``Integrated Perturbation Theory" that in the large-scale limit, the bias shift due to non-Gaussianity in the PS formalism is given by
%%% forget about b_2
\be \Delta b\sub{PS}(k,M)&=&{1\over2} \delta_c (b\sub{PS}-1)\mc{I}(k,M)+{1\over2} \diff{\mc{I}}{\ln\sigma_M}.\lab{delb1}\ee
The function $\mc{I}(k,M)$ is derived from the bispectrum shapes as follows
\be \mc{I}(k,M)={4\fnl\over \mc{M}(k)} \times
\begin{cases}
\displaystyle 1& \mbox{(local)}\\
\displaystyle 3 k^2\bkts{k^{2\alpha_S} \gamma_{2+2\alpha_S}- (1+\alpha_S)^2{\gamma_2 \over 3} }& \mbox{(equilateral)}\\

\displaystyle 3k \gamma_1/2 & \mbox{(folded)}\\

\displaystyle -3k\gamma_1 & \mbox{(orthogonal)}
\end{cases}
\ee
%\be \Delta b_x(k)={\fnl \delta_c b_x(\fnl=0)\over \mc{M}(k)}\times
%\begin{cases}
%2 & \mbox{(local)}\\
%4k^2\gamma_2 & \mbox{(equilateral)}\\
%3k\gamma_1& \mbox{(folded)}\\
%-6k\gamma_1& \mbox{(orthogonal)}
%\end{cases}
%\ee
where 
\be \alpha_S &=&{1-n_s\over 3},\\
\gamma_t (M) &=& {1\over 2\pi^2\sigma_M^2}\int_0^\infty {dk\over k} k^{3-t} W^2(kR)P_L(k).
\nn \ee
Here $W(kR)$ is the Fourier transform of the top-hat window function associated with mass $M=4\pi \rho_0R^3/3$. 
%%%%%It was also shown in \cite{matsubara} that, compared with exact expressions, these large-scale expressions are accurate on scales up to $k\sim0.1 h^{-1}$Mpc. 

%In this approximation, the Eulerian biases for the PS and ST formalisms are given by \re{bLps} and \re{bLst} respectively.

%Those for the Sheth-Tormen mass function read
%\be b_1^L &=& {1\over \delta_c}\bkt{q\nu^2 -1+{2p\over 1+(q\nu^2)^p}},\lab{bLst}\\  
%b_2^L &=&{1\over \delta_c^2}\bkt{q^2\nu^4 -3q\nu^2+{2p(2q\nu^2+2p-1)\over 1+(q\nu^2)^p}}.
%\ee 

%\be \Delta b= {\sigma_M^2\over 2\delta_c^2}\bkts{(2+2\delta_c b_1^L +\delta_c^2b_2^L)\mc{I}(k) +(1+\delta_c b_1^L)\diff{\mc{I}(k)}{\ln\sigma_M} }\ee
%where $b_1^L$ and $b_2^L$ depend on the mass function used and

The main idea is to replace the bias $\bar{b}$ in the Alvarez bias \re{alva} by the non-Gaussian version, which we calculate as follows. First, we substitute the Gaussian bias by the non-Gaussian expression 
\be b\sub{PS} \longrightarrow  b\sub{PS}\super{NG}\equiv b\sub{PS}+\Delta b\sub{PS}(k,m),\ee
where $\Delta b\sub{PS}$ is given by the Matsubara expression \re{delb1}. This allows us to calculate the mass-averaged bias by an averaging analogous to \re{ioi}:
\be \overline{b}(k) \equiv {\int_{M\sub{min}}^\infty b\sub{PS}\super{NG}(k,M) { dn \over d\ln M} dM\over \int_{M\sub{min}}^\infty {dn\over d\ln M} dM}\lab{newbbar},\ee where we have transformed the $d\nu$ integral to an equivalent one in $dM$. The mass function, $dn\over d\ln M$, satisfies
\be{dn\over d \ln M}={\rho_m\over M} f(\nu) {d\ln \sigma^{-1}\over d\ln M}=\rho_m {f(\nu)\over \nu} \diff{\nu}{M}.\ee
However, the mass function is also affected by the presence of non-Gaussianity. Here, we modify the mass function using the prescription of \cite{mvj}, in which the Gaussian mass function is multiplied by a  correction factor $\mc{R}\sub{NG}$:
\be{dn\over d \ln M}&=&\mathcal{R}\sub{NG}{dn\over d \ln M}\bigg|\sub{Gaussian} \\
\mc{R}\sub{NG}&=& \exp\bkt{S_3 \delta_c^3\over 6\sigma_M^2}\bkts{{\delta_c^2\over 6\Delta}\cdot{dS_3\over d\ln \sigma_M} + \Delta},\\
\Delta&\equiv& \sqrt{1-{\delta_c S_3\over3}},\lab{mvj}\ee
where the third cumulant, $S_3(M)$, satisfies the relation (see \eg \cite{me8,desjacques})
\be\sigma^4 S_3(M)&=&{1\over 8\pi^4}\int_0^1 k_1^2 d k_1 \int_0^1 k_2^2d k_2 \int_{-1}^1 d\mu \ff B_\phi(k_1,k_2,k_3)\prod_{i=1}^{3}\mc{A}(k_i,R) ,\lab{mvj}\\
k_3&=&\bkt{k_1^2+k_2^2+2\mu k_1k_2}^{1/2},\qquad  \mc{A}(k_i,R)\equiv \mc{M}(k_i)W(k_iR).\ee 

Putting all these ingredients together, we then obtain the (scale-dependent) non-Gaussian bias in the ionized fraction
\be b_x\super{NG} (k) = {\bar{x}_i -1 \over \bar{x}_i} \ln(1-\bar{x}_i)\bkt{\bar{b}(k)-2}\lab{alvaNG}.\ee

Figure \ref{quad} shows the ionized-fraction bias shift as fraction of the Gaussian bias,
\be r_x\equiv{\Delta b_x\super{NG} - b_x\super{Gaussian}\over b_x\super{Gaussian}},\ee with $r_x=0$ when $\fnl=0$.  The figure shows the results for various shapes of non-Gaussianity with $\fnl=50$, $\zeta_0=50$, evaluated at $z=8$. The curve for the orthogonal type is always negative and its absolute value is shown here. These curves are reminiscent of those for the scale-dependent halo bias seen by previous authors \cite{matsubara, wagner,verde}.  The strongest scale-dependence is exhibited by the local type. The folded type exhibits not only a weak scale-dependence of the bias, but also a low amplitude of the bias shift, and thus we can expect the poorest constraints for the folded-type non-Gaussianity.

\begin{figure}[t]
\centering

\includegraphics[width = 2.5in, angle = -90]{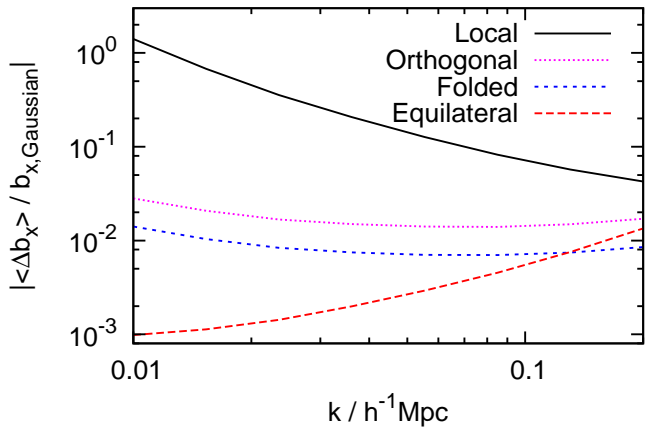}
\caption{The fractional change in the ionized-fraction bias $r_x\equiv\overline{\Delta b_x}/b_{x,\mbox{\scriptsize{Gaussian}}}$ due to non-Gaussianity of various shapes, calculated at $z=8$. The parameters in panel (A) are $\fnl=50$, ionization efficiency $\zeta_0=50$.}
\lab{quad}
\end{figure}

We note that our approach is different from those in \cite{joudaki,tashiro}, which treated the bias-shift in the ionized fraction as an extrapolation of the formula by Dalal \etal \cite{dalal}
\be \Delta b\sub{x}\approx{3\fnl\delta_B(b_x-1)H_0^2\Omega_m \over k^2 T(k) D(z)}, \ee
where $\delta_B\approx1$ is the critical density for collapse in an ionized region. This extension effectively assumes that there is a single physical mechanism governing the clustering of linear density fluctuations and that of ionized hydrogen clouds, even in the presence of primordial non-Gaussianity. This assumption demands further tests from $N-$body simulations with non-Gaussian initial conditions, which have yet to be performed on such scales, nor has it been verified for other types of non-Gaussianity.  Our calculations do not rely on this extrapolation, and is therefore more self-consistent. 

%In later work we shall compare the forecasts on $\fnl$ using our approach and that using such an extrapolation.

%CALCULATION STEPS TO GET 21CM POWER SPECTRUM

%\begin{enumerate}
%\item Choose either PS or ST approach, select shape of NG and value of \fnl.
%\item Use result of $b^L$ and write Eulerian bias as $b^E\approx1+b_L$
%\item Find $\bar{b}$ by averaging $b^E$ over the multiplicity function \re{ioi}.
%\item Choose mean ionization efficiency $\zeta_0$. Find mean fraction $\bar{x}_i$ using \re{xbar}.
%\item Find the ionization bias $b_x$ by using the Alvarez reionization model \re{alva}
%\item Find $\Delta b(k,M)$ by using \re{delb1}-\re{delb2}
%, but replacing $b^L$ with $b_x -1$.
%\item Find mass-average bias shift $\overline{\Delta b}(k)$ using \re{deltabar}
%\item Transform $\overline{\Delta b}$ to $\overline{\Delta b}_x$ by the replacements \re{replace} in the previous two steps.
%\item Finally $\ms{P}_{xx}=(b_x+\Delta b_x)^2 \ms{P}_{\delta\delta}$ etc.
%\end{enumerate}

%Include recent excursion set work by 
%- d'Aloisio (1206.3305) - 2nd order corrections to $f\sub{coll}$ is small for small mass objects.

\section{Fisher-matrix analysis}\lab{fishsec}

We now perform a simple Fisher-matrix analysis to assess the potential of upcoming radio arrays such as the SKA in constraining non-Gaussianity using the 21cm power spectrum.

%Our treatment of noise  follows that \cite{mcquinn} and \cite{maotegmark}, which we outline below.

Radio interferometers do not measure the power spectrum directly in $\mb{k}$ space. Instead, they measure the complex visibility function, $V$, in array-configuration space $\mb{u}=(u,v,u_\parallel)$, where the $(u,v)$ plane is perpendicular to the line of sight and the values of $u$ and $v$ depend on the geometry of the array. In the small-angle approximation, we can write (see \eg \cite{radiobook})
\be V(u,v)=\int d\mb{\hat{n}}\ff \delta T_b(\mb{\hat n})\mc{A}(\mb{\hat{n}})\exp\bkt{2\pi i \begin{pmatrix}u\\v
\end{pmatrix}\cdot \mb{\hat{n}}},\lab{visi}\ee
where $\mc{A}$ is the relative antenna area (equals unity in the line-of-sight direction $\mb{\hat{n}}$). Eq. \ref{visi} can be thought of as the Fourier transform of the 21cm differential brightness temperature (weighted by $\mc{A}$) into the configuration space. 

In the $(u,v)$ plane, we can write $\mb{u}_\perp=\mb{b}/\lambda_{21}(z)$, where $\mb{b}$ is a baseline vector between two receivers. Hence, $\mb{u}_\perp/2\pi$ equals the number of wavelengths that fit between two receivers in the $u$ and $v$ directions. The component $u_\parallel$ can be calculated from the shift in frequency of the signal from the central redshift of a particular redshift bin.

%An observer cannot measure the position vector $\mb{r}$ of the source, but instead, its sky coordinates $\mb{\Theta}$. Let $\mb{r}$ and $\mb{\Theta}$ be represented by $\mb{k}$ and $\mb{u}$ in Fourier space.

The components of $\mb{u}$ along and perpendicular to the line of sight are related to those of the Fourier wavevector $\mb{k}$ by 
\be(\mb{u}_\perp,u_\parallel)=(d_A(z)\mb{k}_\perp, y(z) k_\parallel),\ee
where $d_A$ is the angular diameter distance to redshift $z$ and $y(z)=21\mbox{cm}(1+z)^2/H(z)$. In terms of the baseline $L$, $u_\perp=2\pi L/\lambda (z)$, $\lambda_{21}(z)=21(1+z)$ cm. Using these relations, we can deduce that the power spectrum in $\mb{u}$ space is given by
\be P_{\delta T} (\mb{u}) = {1\over d_A^2 y}P_{\delta T}(\mb k)\lab{poweru}.\ee

The noise power spectrum is given by \cite{mcquinn, maotegmark}
\be P_N(u_\perp)=\bkt{\lambda^2 T\sub{sys}\over A_e}^2{1\over t_0 n(u_\perp)},\lab{noise}\ee
where the system temperature $T\sub{sys}=280((1+z)/7.4)^{2.3}$ K, $t_0$ is the observation time, $A_e$ is the effective collecting area and $n(u_\perp)$ is the number density of baselines that can observe mode $\mb{u}$. For surveys with outer core radius $R_c$ and inner nucleus radius $R_n$, the baseline number density can be expressed as
\be n(u_\perp)&=&2\pi \lambda_{21}^2\int_0^{R_c}rP(r+{\lambda_{21}u_\perp\over 2\pi})P(r)dr,\\
P(r) &=&\begin{cases}
K& 0\leq r< R_n,\\
K\bkt{R_N/r}^2&R_n\leq r <R_c,\\
0 &\mbox{otherwise}.
\end{cases} 
\ee
with normalization constant $K=4e^{1/4}N\sub{ant}/(5\pi R_c^2)$.

%(((
%We estimate this as
%\be n(u_\perp)\approx 
%\begin{cases}
%%%%%{1\over2 A_e}\bkt{N\sub{ant}\lambda_{21}\over 2\pi}^2, & \mbox{if }L\sub{min}< \lambda_{21}u_\perp/2\pi< R\sub{c},\\
%50 \mbox{ (SKA) }/3\times10^6 \mbox{ (Futuristic)} & \mbox{if }L\sub{min}< \lambda_{21}u_\perp/2\pi< R\sub{c},\\

%0 &\mbox{ otherwise,} 
%\end{cases}
%\lab{nperp}\ee

%)))

\begin{table}
\begin{center}
\begin{tabular}{|c|c|c|c|c|c|} 
\hline
& $N\sub{ant}$ & $L\sub{min}$ (m) & $R\sub{c}$ (m)  & FoV  (deg$^2$)) & $A_e$ (m$^2$), $z=8/10/12$ \\
\hline
SKA  & 1400 &10  &  330 &  232 &50/77/104\\ 
Futuristic & $10^6$ & 1 & 1000 & 20626 & 1/1/1 \\
\hline 
\end{tabular}
\end{center}
\caption{Parameters adopted for the Fisher forecasts for the SKA and a futuristic radio array.}
\lab{tab}
\end{table}

The error in the measurement in each pixel is given by
\be \delta P(\mb{u})={P_{\delta T}(\mb{u})+ P_N(u_\perp) \over \sqrt{N}}\lab{delpu}\ee
where 
\be N = \bkt{k\over2\pi}^2 \sin\theta \ff V(z)  \Delta\theta\Delta k, \lab{Ncell}\ee
is the number of cells in an annulus defined by $\Delta k$ and $\Delta\theta$ spanning half the sphere, and $V(z)=d_A^2 yB\times$FoV, where FoV is the area of the field of view. $B$ (sometimes called the bandwidth) is related to the line-of-sight distance, $\Delta r$,  from the central redshift of a particular bin by $B=\Delta r/y(z)$.

Finally, the Fisher matrix is given by
\be F_{\alpha\beta}=\sum{1\over [\delta P (\mb{u})]^2}\pdiff{P_{\delta T}(\mb{u})}{p_\alpha}\pdiff{P_{\delta T}(\mb{u})}{p_\beta},\lab{fish}\ee
where the summation extends over the valid pixels in $\mb{k}$ space. This assumes a Gaussian uncertainty in each parameter $p$ (see \cite{norena} for an alternative treatment). Assuming further that the noise correlation across redshift bins can be neglected, we can add the Fisher matrices evaluated at various redshifts.  

%%%% IMPORTANT FOR CODE:::
%%% Note that for numerical purposes.. use  \re{shortcut} the fnl derivative can be evaluated as 
%%$$ \pdiff{P_{\delta T}(\mb u)}{\fnl} ={r_x \mc{P}_{\delta\delta}(\mb k) b_x(b_x\super{NG}+b_x\super{Gaussian} -2 -2\mu^2)\over \Delta\fnl (d_A^2 y)}$$

\begin{table}
\begin{center}
\begin{tabular}{|c|c|c|} 
\hline
Template & $\sigma(\fnl)$ SKA   &$\sigma(\fnl)$ Fut. \\
\hline % sigma(zeta) 
% top entries use B= 6, bottom ones use z-dependent values.
%Local &  53 & 8& 4 & 0.5 \\  %0.13,  0.23, 0.01,0.015
%Equilateral & 181& 62 & 13   &  4 \\ %0.09,  0.19,0.007, 0.012
%Folded & 309& 122 & 30 & 10\\ %0.11 , 0.32, 0.011, 0.025
%Orthogonal & 154 & 61 &  15 & 6\\ %0.11 , 0.32 ,0.011, 0.03
Local &  2 &  0.1 \\  %0.13,  0.23, 0.01,0.015
Equilateral & 13 & 1  \\ %0.09,  0.19,0.007, 0.012
Folded & 30 & 3 \\ %0.11 , 0.32, 0.011, 0.025
Orthogonal & 15 &   1 \\ %0.11 , 0.32 ,0.011, 0.03
\hline 
\end{tabular}
\end{center}
\caption{Forecasts on the $\fnl$ constraints of various shapes using the \ii{SKA} and \ii{Futuristic} specifications in Table \ref{tab}, and assuming the Press-Schechter collapse formalism.}
\lab{tabfnl}
\end{table}

In this work, we shall consider two setups, one for the SKA and another for a futuristic radio array (such as one considered in \cite{maotegmark, lazio}). The specifications for these setups are given in Table \ref{tab}. We consider a simple parameter space consisting of the non-Gaussianity amplitude and the ionization efficiency, \ie, $p_\alpha=(\fnl,\zeta_0)$. The background cosmology is fixed using the best-fit parameters outlined in the Introduction. Whilst including more parameters in the Fisher forecast would most likely yield less stringent constraints on these parameters, previous studies taking a similar approach have estimated that when including a full set of cosmological parameters, the local $\fnl$ constraints would be affected only by $\sim10\%$ \cite{joudaki, norena}. Thus we believe that the forecasts we obtain here are indicative of the constraints future radio arrays would be able to achieve.

%(This differs from the analysis in \cite{joudaki}, who treated the reionization-fraction bias $b_x$ as an additional free parameter.) 

We proceed to calculate the Fisher matrix as follows. Firstly, a set of experimental specifications is chosen (see Table \ref{tab}) assuming a fixed redshift  ($z=8/10/12$). We set the fiducial values of the parameters to be $\zeta_0=50$, $\fnl=0$ and calculate the Fisher matrix over the range $k\in[k\sub{min},k\sub{max}]$ and $\theta\in[\theta\sub{min},\theta\sub{max}]$. The value of $k\sub{min}=2\pi/yB$ is set by the largest scale beyond which the signal is dominated by foregrounds (with $B=8.3/ 7/ 7.5$ MHz), and $k\sub{max}=0.2$ Mpc  determines the shortest wavelength below which linear calculations break down (see \cite{maotegmark,joudaki} for studies of the variation in $k\sub{max}$). For a given $k$, we determine the range of $\theta$ such that the mode can be observed by the survey, \iee those values of $\theta$ such that $u_\perp = d_A k\sin\theta$.
% we only need to scan $\theta$ in the range
%\be \mbox{max}\left\{0,\sin^{-1}\bkt{u_{\perp,\mbox{\scriptsize{min}}}\over d_Ak}\right\}\leq \theta \leq  
%\mbox{min}\left\{{\pi\over 2},\sin^{-1}\bkt{u_{\perp,\mbox{\scriptsize{max}}}\over d_A k}\right\}
%\ee

%In these bins, we set $B=8.3/7.5/7$ MHz respectively. 
% In code set B = 6 throughout.

%\begin{enumerate}
%Pick resolution $\Delta k$. 
%\item Also choose the angular resolution $\Delta \theta$.
%\item For this value of $k$, find $P_{\delta T}(\mb{u})$ using \re{power}  and \re{poweru}.
%\item Find the corresponding $u_\perp$ and hence the noise \re{noise}, and number of modes in each cell \re{Ncell}.
%\item Get error in this pixel \re{delpu}.
%\end{enumerate}

The Fisher forecasts on $\fnl$ of various shapes are shown in Table \ref{tabfnl}. Our constraint on the local shape is $\sigma(\fnl)= [2,0.1]$ for the [SKA,Futuristic] configurations. The SKA constraint in comparable with those expected from Planck. The equilateral and orthogonal constraints are similar in both the amplitudes and the scaling and thus distinguishing between these shapes using the power spectrum will be challenging. The folded orthognal fared worst as expected, with the error roughly twice that of the orthogonal shape. For the latter 3 shapes, the constraints are poorer than the local shape since the region of the strongest signals are on large scales which are dominated by foregrounds. In all cases, the SKA constraints will be improved by at least an order of magnitude with the Futuristic array, down to $\sigma(\fnl)=\mc{O}(1)$ for all shapes.

%hese results are consistent with those of \cite{joudaki} who obtained $\sigma(\fnl)= [50,4]$ (though they did not appeal to a specific reionization model). 

Our results suggest that there are realistic prospects of contraining the shapes of non-Gaussianity with future 21cm power-spectrum measurements.  Although the 21cm constraints on $\fnl$  will not be markedly different from those from upcoming CMB experiments, 21cm experiments are complementary to the latter as they constrain $\fnl$ on a much smaller scale. A multi-scale attack on non-Gaussianity is required if the contributions from the various shapes were to be disentangled, and this can be achieved by combining non-Gaussianity constraints from the CMB, galaxy clusters and 21cm measurements. Such approach will also shed light on the possible scale-dependence of non-Gaussianity \cite{becker,byrnes}.

%% WAIT>>>>>>>>
We also carried out the Fisher-matrix analysis using the Sheth-Tormen formalism (see Appendix \ref{aaa}). The values of $\sigma(\fnl)$, shown in Table \re{tabsheth}, are roughly twice that of the PS formalism. This is in line with expectation from previous studies comparing cluster biases in the PS and ST formalisms \cite{me, shapiro, matsubara}, which found that the amplitude of the bias in the ST formalism is significantly less than that in the PS formalism. Alternatively, one could also extend our calculations to other functions based on more recent $N$-body simulations \cite{warren,reed}, but the bias shift \re{delb1} will be more difficult to obtain analytically. Nevertheless, we expect such calculations to produce $\fnl$ constraints that are compatible with the PS and ST analyses.

Our constraints rely on the assumption that foregrounds can be perfectly subtracted on scales $k>k\sub{min}$. However, even on these scales, temperature fluctuations due to Galactic synchroton, extra-galactic point sources and free-free emissions can still significantly overwhelm the 21cm signal, and its subtraction can remove powers that carry essential information from primordial non-Gaussianity. Foreground cleaning by polynomial fitting (see \eg  \cite{jelic,liu} and references therein) could also possibly ``fit out" the non-Gaussian features, and a non-parametric cleaning scheme \cite{harker, chapman} may be a more appealing alternative. It is hoped that by the time the SKA comes into operation, its precursors \cite{bernardi,ghosh,jelic2} would have shed enough light on the nature of foregrounds that they may be feasibly subtracted.

\begin{table}
\begin{center}
\begin{tabular}{|c|c|c|} 
\hline
Template & $\sigma(\fnl)$ SKA   &$\sigma(\fnl)$ Fut. \\
\hline % sigma(zeta) 
Local &  4 &  0.3 \\  %0.13,  0.23, 0.01,0.015
Equilateral & 23 & 2  \\ %0.09,  0.19,0.007, 0.012
Folded & 56 & 5 \\ %0.11 , 0.32, 0.011, 0.025
Orthogonal & 28 &   2 \\ %0.11 , 0.32 ,0.011, 0.03
\hline 
\end{tabular}
\end{center}
\caption{Forecasts on the $\fnl$ constraints of various shapes using the Sheth-Tormen formalism. }
\lab{tabsheth}
\end{table}

%%%%% MORE emphasis on shapes?

%did not address issue of GR correction on large scale - could mimic fnl local \cite{verde}

\section{Conclusions}

We have shown how information on primordial non-Gaussianity can be extracted from the 21cm radiation signals -- a connection which deserves to be explored more deeply. Our primary focus was to study the manifestation of various shapes of non-Gaussianity in the 21cm power spectrum, and  thus evaluating the prospects for future radio telescopes in detecting these signatures. Our calculations are based on recent analytic work on the bias of large-scale structures, modified and extended to scales where there are good prospects of detecting the 21cm signals from the epoch of reionization.  We found that the shapes of non-Gaussianity manifest as different scale dependence in the bias of ionized fraction as shown in Fig. \ref{quad}. 

Our analysis demonstrated that the $\fnl$ constraints from the SKA will be competitive with those from upcoming CMB experiments such as Planck (and much better than current LSS constraints), whilst futuristic radio arrays can tighten the $\fnl$ constraints down by at least an order of magnitude towards $\sigma(\fnl)=\mc{O}(1)$. There are several ways in which the constraints  can be further improved. Firstly, we note that our analysis uses only the information in the power spectrum of the 21cm radiation, whereas much stronger constraints are expected from measurements of the bispectrum \cite{cooray}. Secondly, the 21cm measurements can be made across a larger number of redshift bins. Indeed, 21cm tomography has been shown to potentially outperform CMB experiments (which provide measurements of fluctuations at a single redshift) in constraining key cosmological parameters \cite{maotegmark}. Finally, the CMB, large-scale structures and 21cm signals probe non-Gaussianity on different physical and time scales, and combining these constraints will be crucial in the understanding of complex physics such as scale-dependence and time-evolution of \fnl.

% need more work on simulation and analytical progress from excursion set.

%test sensitivity to $J_\alpha$?.. 

%These are left for future investigations.

%In the optimistic case....detect various shapes... new physics!

A detection of non-Gaussianity will clearly have far-reaching consequence for our understanding of cosmology, and the power of the 21cm signal in probing non-Gaussianity is only becoming better understood within recent years. With future work, we plan to show how the 21cm radiation can be used not only to constrain but also to \ii{disentangle} various shapes of non-Gaussianity. This will be crucial in the understanding of the physics underlying inflation.

%It is most likely, however, that the first constraints will come from measurements of the CMB anisotropies.

%On the other hand, a pessimistic outlook is one in which primordial non-Gaussianity continues to elude detection. ... search continues... 
%However, first constraints will come from the CMB. It may also be possible to detect such non-Gaussianity signal from the 21cm radiation from the first minihaloes \cite{me}.

\acknowledgments 
I am grateful to Shahab Joudaki, Hiroyuki Tashiro and Avery Meiksin for helpful discussions. I thank the referee for suggestions which led to  significant improvement of the initial results.

\appendix

\section[Appendix : Extension to the Sheth-Tormen formalism]{21cm power spectrum with the Sheth-Tormen formalism}\lab{aaa}

% mention zeta = 50 in figure reference.

% NOTE .. "multiplicity" function is off by factor of nu from standard definition.. as integration is actually over ln(nu)

We outline a method of extending our calculation of the ionized-fraction bias $b_x$, based on the Press-Schechter collapse formalism, to the Sheth-Tormen formalism. The idea is based on averaging the large-scale bias over the mass function:
\be \bar{b}\sub{ST} &=&  {\int_{\nu\sub{min}}^\infty f\sub{ST}(\nu) b\sub{ST}(\nu) d\ln\nu\over   \int_{\nu\sub{min}}^\infty f\sub{ST}(\nu) d\ln\nu},\\
b\sub{ST}&=& 1+  {1\over \delta_c}\bkt{q\nu^2 -1+{2p\over 1+(q\nu^2)^p}},\lab{bLst}\quad p=0.3, \ff q=0.707,\\
 f\sub{ST}(\nu)&=& {\half}\nu e^{-{q\nu^2/2}}\bkts{1+\bkt{q\nu^2}^{-p}}.\ee
Note that these equations reduce to the PS case if $p=0$ and $q=1$.

To calculate $b_x$ in our chosen reionization model \re{alva}, we also need the collapse fraction,  $f\sub{coll}$, which, in the Sheth-Tormen formalism is quite complex (see the expressions derived from excursion-set theory in \eg \citep{adshead,daloisio}). However, the scaling $\delta_c\rightarrow \sqrt{q}\delta_c$ was found to be a reasonable approximation for ellipsoidal collapse integral to the Sheth-Tormen approach \citep{giocoli,desimone}.

From \cite{matsubara}, the bias shift is given by

\be \Delta b\sub{ST}(k,M)&=&\bkts{{q\delta_c (b\sub{ST}-1)\over 2}+ {p(q\nu^2+2p+1)\over\nu^2(1+(q\nu^2)^p)}}\mc{I}(k,M)\nn\\
&\ff&+ \bkts{{q\over2} +{p\over \nu^2(1+(q\nu^2)^p)}  } \diff{\mc{I}}{\ln\sigma_M},\lab{delb2}\ee

\no These expressions together with the Alvarez bias \re{alvaNG} yield the ionized-fraction bias, $b_x$.

\bibliographystyle{jhep}
\bibliography{21power}

\providecommand{\href}[2]{#2}\begingroup\raggedright\begin{thebibliography}{10}

\bibitem{maldacena}
J.~{Maldacena}, {\it {Non-gaussian features of primordial fluctuations in
  single field inflationary models}},  {\em JHEP} {\bf 5} (May, 2003) 13,
  [\href{http://xxx.lanl.gov/abs/astro-ph/0210603}{{\tt astro-ph/0210603}}].

\bibitem{chen}
X.~{Chen}, {\it {Primordial Non-Gaussianities from Inflation Models}},  {\em
  Adv. Astron.} {\bf 2010} (2010)
  [\href{http://xxx.lanl.gov/abs/1002.1416}{{\tt arXiv:1002.1416}}].

\bibitem{bartolo}
N.~Bartolo, E.~Komatsu, S.~Matarrese, and A.~Riotto, {\it {Non-Gaussianity from
  inflation: Theory and observations}},  {\em Phys. Rept.} {\bf 402} (2004)
  103--266.

\bibitem{komatsu}
E.~{Komatsu} et~al., {\it {Seven-year Wilkinson Microwave Anisotropy Probe
  (WMAP) Observations: Cosmological Interpretation}},  {\em \apjs} {\bf 192}
  (Feb., 2011) 18, [\href{http://xxx.lanl.gov/abs/1001.4538}{{\tt
  arXiv:1001.4538}}].

\bibitem{creminelli}
P.~{Creminelli}, A.~{Nicolis}, L.~{Senatore}, M.~{Tegmark}, and
  M.~{Zaldarriaga}, {\it {Limits on non-Gaussianities from WMAP data}},  {\em
  \jcap} {\bf 5} (May, 2006) 4,
  [\href{http://xxx.lanl.gov/abs/astro-ph/0509029}{{\tt astro-ph/0509029}}].

\bibitem{meerburg}
P.~D. {Meerburg}, J.~P. {van der Schaar}, and P.~{Stefano Corasaniti}, {\it
  {Signatures of initial state modifications on bispectrum statistics}},  {\em
  \jcap} {\bf 5} (May, 2009) 18, [\href{http://xxx.lanl.gov/abs/0901.4044}{{\tt
  arXiv:0901.4044}}].

\bibitem{senatore}
L.~{Senatore}, K.~M. {Smith}, and M.~{Zaldarriaga}, {\it {Non-Gaussianities in
  single field inflation and their optimal limits from the WMAP 5-year data}},
  {\em \jcap} {\bf 1} (Jan., 2010) 28,
  [\href{http://xxx.lanl.gov/abs/0905.3746}{{\tt arXiv:0905.3746}}].

\bibitem{desjacques}
V.~{Desjacques} and U.~{Seljak}, {\it {Primordial non-Gaussianity from the
  large-scale structure}},  {\em Classical and Quantum Gravity} {\bf 27} (June,
  2010) 124011, [\href{http://xxx.lanl.gov/abs/1003.5020}{{\tt
  arXiv:1003.5020}}].

\bibitem{cooray}
A.~{Cooray}, {\it {21-cm Background Anisotropies Can Discern Primordial
  Non-Gaussianity}},  {\em Phys. Rev. Lett.} {\bf 97} (Dec., 2006) 261301,
  [\href{http://xxx.lanl.gov/abs/astro-ph/0610257}{{\tt astro-ph/0610257}}].

\bibitem{joudaki}
S.~{Joudaki}, O.~{Dor{\'e}}, L.~{Ferramacho}, M.~{Kaplinghat}, and M.~G.
  {Santos}, {\it {Primordial Non-Gaussianity from the 21 cm Power Spectrum
  during the Epoch of Reionization}},  {\em Phys. Rev. Lett.} {\bf 107} (Sept.,
  2011) 131304, [\href{http://xxx.lanl.gov/abs/1105.1773}{{\tt
  arXiv:1105.1773}}].

\bibitem{pillepich}
A.~{Pillepich}, C.~{Porciani}, and S.~{Matarrese}, {\it {The Bispectrum of
  Redshifted 21 Centimeter Fluctuations from the Dark Ages}},  {\em \apj} {\bf
  662} (June, 2007) 1--14,
  [\href{http://xxx.lanl.gov/abs/astro-ph/0611126}{{\tt astro-ph/0611126}}].

\bibitem{tashiro}
H.~{Tashiro} and S.~{Ho}, {\it {Constraining primordial non-Gaussianity with
  CMB-21cm cross-correlations?}},  {\em ArXiv: 1205.0563} (May, 2012)
  [\href{http://xxx.lanl.gov/abs/1205.0563}{{\tt arXiv:1205.0563}}].

\bibitem{tashiro1}
H.~{Tashiro} and N.~{Sugiyama}, {\it {Ionized bubble number count as a probe of
  non-Gaussianity}},  {\em \mnras} {\bf 420} (Feb., 2012) 441--446,
  [\href{http://xxx.lanl.gov/abs/1104.0149}{{\tt arXiv:1104.0149}}].

\bibitem{me}
S.~{Chongchitnan} and J.~{Silk}, {\it {The 21-cm radiation from minihaloes as a
  probe of small primordial non-Gaussianity}},  {\em \mnras} (Aug., 2012) L507,
  [\href{http://xxx.lanl.gov/abs/1205.6799}{{\tt arXiv:1205.6799}}].

\bibitem{lahavliddle}
O.~{Lahav} and A.~R. {Liddle}, {\it {The Cosmological Parameters 2010}},  {\em
  ArXiv e-prints} (Feb., 2010) [\href{http://xxx.lanl.gov/abs/1002.3488}{{\tt
  arXiv:1002.3488}}].

\bibitem{gangui}
A.~{Gangui}, F.~{Lucchin}, S.~{Matarrese}, and S.~{Mollerach}, {\it {The
  three-point correlation function of the cosmic microwave background in
  inflationary models}},  {\em \apj} {\bf 430} (Aug., 1994) 447--457,
  [\href{http://xxx.lanl.gov/abs/astro-ph/9312033}{{\tt astro-ph/9312033}}].

\bibitem{verdewang}
L.~{Verde}, L.~{Wang}, A.~F. {Heavens}, and M.~{Kamionkowski}, {\it
  {Large-scale structure, the cosmic microwave background and primordial
  non-Gaussianity}},  {\em \mnras} {\bf 313} (Mar., 2000) 141--147,
  [\href{http://xxx.lanl.gov/abs/astro-ph/9906301}{{\tt astro-ph/9906301}}].

\bibitem{komatsuspergel}
E.~{Komatsu} and D.~N. {Spergel}, {\it {Acoustic signatures in the primary
  microwave background bispectrum}},  {\em \prd} {\bf 63} (Mar., 2001) 063002,
  [\href{http://xxx.lanl.gov/abs/astro-ph/0005036}{{\tt astro-ph/0005036}}].

\bibitem{wagner}
C.~{Wagner} and L.~{Verde}, {\it {N-body simulations with generic non-Gaussian
  initial conditions II: halo bias}},  {\em \jcap} {\bf 3} (Mar., 2012) 2,
  [\href{http://xxx.lanl.gov/abs/1102.3229}{{\tt arXiv:1102.3229}}].

\bibitem{alvarez}
M.~A. {Alvarez}, E.~{Komatsu}, O.~{Dor{\'e}}, and P.~R. {Shapiro}, {\it {The
  Cosmic Reionization History as Revealed by the Cosmic Microwave Background
  Doppler-21 cm Correlation}},  {\em \apj} {\bf 647} (Aug., 2006) 840--852,
  [\href{http://xxx.lanl.gov/abs/astro-ph/0512010}{{\tt astro-ph/0512010}}].

\bibitem{bcek}
J.~R. {Bond}, S.~{Cole}, G.~{Efstathiou}, and N.~{Kaiser}, {\it {Excursion set
  mass functions for hierarchical Gaussian fluctuations}},  {\em \apj} {\bf
  379} (Oct., 1991) 440--460.

\bibitem{lacey}
C.~{Lacey} and S.~{Cole}, {\it {Merger rates in hierarchical models of galaxy
  formation}},  {\em \mnras} {\bf 262} (June, 1993) 627--649.

\bibitem{mowhite}
H.~J. {Mo} and S.~D.~M. {White}, {\it {An analytic model for the spatial
  clustering of dark matter haloes}},  {\em \mnras} {\bf 282} (Sept., 1996)
  347--361, [\href{http://xxx.lanl.gov/abs/astro-ph/9512127}{{\tt
  astro-ph/9512127}}].

\bibitem{sheth}
R.~K. {Sheth} and G.~{Tormen}, {\it {An excursion set model of hierarchical
  clustering: ellipsoidal collapse and the moving barrier}},  {\em \mnras} {\bf
  329} (Jan., 2002) 61--75,
  [\href{http://xxx.lanl.gov/abs/astro-ph/0105113}{{\tt astro-ph/0105113}}].

\bibitem{zaldarriaga}
M.~{Zaldarriaga}, S.~R. {Furlanetto}, and L.~{Hernquist}, {\it {21 Centimeter
  Fluctuations from Cosmic Gas at High Redshifts}},  {\em \apj} {\bf 608}
  (June, 2004) 622--635, [\href{http://xxx.lanl.gov/abs/astro-ph/0311514}{{\tt
  astro-ph/0311514}}].

\bibitem{bharadwaj}
S.~{Bharadwaj} and S.~S. {Ali}, {\it {On using visibility correlations to probe
  the HI distribution from the dark ages to the present epoch - I. Formalism
  and the expected signal}},  {\em \mnras} {\bf 356} (Feb., 2005) 1519--1528,
  [\href{http://xxx.lanl.gov/abs/astro-ph/0406676}{{\tt astro-ph/0406676}}].

\bibitem{barkana}
R.~{Barkana} and A.~{Loeb}, {\it {A Method for Separating the Physics from the
  Astrophysics of High-Redshift 21 Centimeter Fluctuations}},  {\em \apjl} {\bf
  624} (May, 2005) L65--L68,
  [\href{http://xxx.lanl.gov/abs/astro-ph/0409572}{{\tt astro-ph/0409572}}].

\bibitem{dalal}
N.~Dalal, O.~Dore, D.~Huterer, and A.~Shirokov, {\it {The imprints of
  primordial non-gaussianities on large- scale structure: scale dependent bias
  and abundance of virialized objects}},  {\em Phys. Rev.} {\bf D77} (2008)
  123514.

\bibitem{matsubara}
T.~{Matsubara}, {\it {Deriving an accurate formula of scale-dependent bias with
  primordial non-Gaussianity: An application of the integrated perturbation
  theory}},  {\em \prd} {\bf 86} (Sept., 2012) 063518,
  [\href{http://xxx.lanl.gov/abs/1206.0562}{{\tt arXiv:1206.0562}}].

\bibitem{verde}
L.~{Verde} and S.~{Matarrese}, {\it {Detectability of the Effect of
  Inflationary Non-Gaussianity on Halo Bias}},  {\em \apjl} {\bf 706} (Nov.,
  2009) L91--L95, [\href{http://xxx.lanl.gov/abs/0909.3224}{{\tt
  arXiv:0909.3224}}].

\bibitem{mvj}
S.~{Matarrese}, L.~{Verde}, and R.~{Jimenez}, {\it {The Abundance of
  High-Redshift Objects as a Probe of Non-Gaussian Initial Conditions}},  {\em
  \apj} {\bf 541} (Sept., 2000) 10--24.

\bibitem{me8}
S.~{Chongchitnan} and J.~{Silk}, {\it {A Study of High-order Non-Gaussianity
  with Applications to Massive Clusters and Large Voids}},  {\em \apj} {\bf
  724} (Nov., 2010) 285--295.

\bibitem{radiobook}
B.~F. {Burke} and F.~{Graham-Smith}, {\em An introduction to radio astronomy}.
\newblock Cambridge University Press, Cambridge, 3rd~ed., 2010.

\bibitem{mcquinn}
M.~{McQuinn}, O.~{Zahn}, M.~{Zaldarriaga}, L.~{Hernquist}, and S.~R.
  {Furlanetto}, {\it {Cosmological Parameter Estimation Using 21 cm Radiation
  from the Epoch of Reionization}},  {\em \apj} {\bf 653} (Dec., 2006)
  815--834, [\href{http://xxx.lanl.gov/abs/astro-ph/0512263}{{\tt
  astro-ph/0512263}}].

\bibitem{maotegmark}
Y.~{Mao}, M.~{Tegmark}, M.~{McQuinn}, M.~{Zaldarriaga}, and O.~{Zahn}, {\it
  {How accurately can 21cm tomography constrain cosmology?}},  {\em \prd} {\bf
  78} (July, 2008) 023529, [\href{http://xxx.lanl.gov/abs/0802.1710}{{\tt
  arXiv:0802.1710}}].

\bibitem{norena}
J.~{Nore{\~n}a}, L.~{Verde}, G.~{Barenboim}, and C.~{Bosch}, {\it {Prospects
  for constraining the shape of non-Gaussianity with the scale-dependent
  bias}},  {\em \jcap} {\bf 8} (Aug., 2012) 19,
  [\href{http://xxx.lanl.gov/abs/1204.6324}{{\tt arXiv:1204.6324}}].

\bibitem{lazio}
J.~Lazio, C.~Carilli, J.~Hewitt, S.~Furlanetto, and J.~Burns, {\it The lunar
  radio array (lra)}, .

\bibitem{becker}
A.~{Becker} and D.~{Huterer}, {\it {First Constraints on the Running of
  Non-Gaussianity}},  {\em Physical Review Letters} {\bf 109} (Sept., 2012)
  121302, [\href{http://xxx.lanl.gov/abs/1207.5788}{{\tt arXiv:1207.5788}}].

\bibitem{byrnes}
C.~T. {Byrnes}, M.~{Gerstenlauer}, S.~{Nurmi}, G.~{Tasinato}, and D.~{Wands},
  {\it {Scale-dependent non-Gaussianity probes inflationary physics}},  {\em
  \jcap} {\bf 10} (Oct., 2010) 4,
  [\href{http://xxx.lanl.gov/abs/1007.4277}{{\tt arXiv:1007.4277}}].

\bibitem{shapiro}
P.~R. {Shapiro}, K.~{Ahn}, M.~A. {Alvarez}, I.~T. {Iliev}, H.~{Martel}, and
  D.~{Ryu}, {\it {The 21 cm Background from the Cosmic Dark Ages: Minihalos and
  the Intergalactic Medium before Reionization}},  {\em \apj} {\bf 646} (Aug.,
  2006) 681--690, [\href{http://xxx.lanl.gov/abs/astro-ph/0512516}{{\tt
  astro-ph/0512516}}].

\bibitem{warren}
M.~S. {Warren}, K.~{Abazajian}, D.~E. {Holz}, and L.~{Teodoro}, {\it {Precision
  Determination of the Mass Function of Dark Matter Halos}},  {\em \apj} {\bf
  646} (Aug., 2006) 881--885,
  [\href{http://xxx.lanl.gov/abs/astro-ph/0506395}{{\tt astro-ph/0506395}}].

\bibitem{reed}
D.~S. {Reed}, R.~E. {Smith}, D.~{Potter}, A.~{Schneider}, J.~{Stadel}, and
  B.~{Moore}, {\it {Toward an accurate mass function for precision cosmology}},
   {\em ArXiv e-prints} (June, 2012)
  [\href{http://xxx.lanl.gov/abs/1206.5302}{{\tt arXiv:1206.5302}}].

\bibitem{jelic}
V.~{Jeli{\'c}}, S.~{Zaroubi}, P.~{Labropoulos}, R.~M. {Thomas}, G.~{Bernardi},
  M.~A. {Brentjens}, A.~G. {de Bruyn}, B.~{Ciardi}, G.~{Harker}, L.~V.~E.
  {Koopmans}, V.~N. {Pandey}, J.~{Schaye}, and S.~{Yatawatta}, {\it {Foreground
  simulations for the LOFAR-epoch of reionization experiment}},  {\em \mnras}
  {\bf 389} (Sept., 2008) 1319--1335,
  [\href{http://xxx.lanl.gov/abs/0804.1130}{{\tt arXiv:0804.1130}}].

\bibitem{liu}
A.~{Liu}, M.~{Tegmark}, J.~{Bowman}, J.~{Hewitt}, and M.~{Zaldarriaga}, {\it
  {An improved method for 21-cm foreground removal}},  {\em \mnras} {\bf 398}
  (Sept., 2009) 401--406, [\href{http://xxx.lanl.gov/abs/0903.4890}{{\tt
  arXiv:0903.4890}}].

\bibitem{harker}
G.~{Harker}, S.~{Zaroubi}, G.~{Bernardi}, M.~A. {Brentjens}, A.~G. {de Bruyn},
  B.~{Ciardi}, V.~{Jeli{\'c}}, L.~V.~E. {Koopmans}, P.~{Labropoulos},
  G.~{Mellema}, A.~{Offringa}, V.~N. {Pandey}, J.~{Schaye}, R.~M. {Thomas}, and
  S.~{Yatawatta}, {\it {Non-parametric foreground subtraction for 21-cm epoch
  of reionization experiments}},  {\em \mnras} {\bf 397} (Aug., 2009)
  1138--1152, [\href{http://xxx.lanl.gov/abs/0903.2760}{{\tt
  arXiv:0903.2760}}].

\bibitem{chapman}
E.~{Chapman}, F.~B. {Abdalla}, G.~{Harker}, V.~{Jeli{\'c}}, P.~{Labropoulos},
  S.~{Zaroubi}, M.~A. {Brentjens}, A.~G. {de Bruyn}, and L.~V.~E. {Koopmans},
  {\it {Foreground removal using FASTICA: a showcase of LOFAR-EoR}},  {\em
  \mnras} {\bf 423} (July, 2012) 2518--2532,
  [\href{http://xxx.lanl.gov/abs/1201.2190}{{\tt arXiv:1201.2190}}].

\bibitem{bernardi}
G.~{Bernardi}, A.~G. {de Bruyn}, M.~A. {Brentjens}, B.~{Ciardi}, G.~{Harker},
  V.~{Jeli{\'c}}, L.~V.~E. {Koopmans}, P.~{Labropoulos}, A.~{Offringa}, V.~N.
  {Pandey}, J.~{Schaye}, R.~M. {Thomas}, S.~{Yatawatta}, and S.~{Zaroubi}, {\it
  {Foregrounds for observations of the cosmological 21 cm line. I. First
  Westerbork measurements of Galactic emission at 150 MHz in a low latitude
  field}},  {\em \aap} {\bf 500} (June, 2009) 965--979,
  [\href{http://xxx.lanl.gov/abs/0904.0404}{{\tt arXiv:0904.0404}}].

\bibitem{ghosh}
A.~{Ghosh}, J.~{Prasad}, S.~{Bharadwaj}, S.~S. {Ali}, and J.~N. {Chengalur},
  {\it {Characterizing foreground for redshifted 21 cm radiation: 150 MHz Giant
  Metrewave Radio Telescope observations}},  {\em \mnras} {\bf 426} (Nov.,
  2012) 3295--3314, [\href{http://xxx.lanl.gov/abs/1208.1617}{{\tt
  arXiv:1208.1617}}].

\bibitem{jelic2}
V.~{Jeli{\'c}}, S.~{Zaroubi}, P.~{Labropoulos}, G.~{Bernardi}, A.~G. {de
  Bruyn}, and L.~V.~E. {Koopmans}, {\it {Realistic simulations of the Galactic
  polarized foreground: consequences for 21-cm reionization detection
  experiments}},  {\em \mnras} {\bf 409} (Dec., 2010) 1647--1659,
  [\href{http://xxx.lanl.gov/abs/1007.4135}{{\tt arXiv:1007.4135}}].

\bibitem{adshead}
P.~{Adshead}, E.~J. {Baxter}, S.~{Dodelson}, and A.~{Lidz}, {\it
  {Non-Gaussianity and excursion set theory: Halo bias}},  {\em \prd} {\bf 86}
  (Sept., 2012) 063526, [\href{http://xxx.lanl.gov/abs/1206.3306}{{\tt
  arXiv:1206.3306}}].

\bibitem{daloisio}
A.~{D'Aloisio}, J.~{Zhang}, D.~{Jeong}, and P.~R. {Shapiro}, {\it {Halo
  statistics in non-Gaussian cosmologies: the collapsed fraction, conditional
  mass function, and halo bias from the path-integral excursion set method}},
  {\em ArXiv e-prints} (June, 2012)
  [\href{http://xxx.lanl.gov/abs/1206.3305}{{\tt arXiv:1206.3305}}].

\bibitem{giocoli}
C.~{Giocoli}, J.~{Moreno}, R.~K. {Sheth}, and G.~{Tormen}, {\it {An improved
  model for the formation times of dark matter haloes}},  {\em \mnras} {\bf
  376} (Apr., 2007) 977--983,
  [\href{http://xxx.lanl.gov/abs/astro-ph/0611221}{{\tt astro-ph/0611221}}].

\bibitem{desimone}
A.~{de Simone}, M.~{Maggiore}, and A.~{Riotto}, {\it {Conditional probabilities
  in the excursion set theory: generic barriers and non-Gaussian initial
  conditions}},  {\em \mnras} {\bf 418} (Dec., 2011) 2403--2421,
  [\href{http://xxx.lanl.gov/abs/1102.0046}{{\tt arXiv:1102.0046}}].

\end{thebibliography}\endgroup

\end{document}